\documentclass[a4paper,twocolumn, altadd, nofootinbib,superscriptaddress,aps,prb,citeautoscript, 14pt,
eqsecnum,notitlepage,showkeys]{revtex4-1}
\usepackage{multirow}
\usepackage{graphicx}
\usepackage{bm}
\usepackage{dcolumn}
\usepackage[hidelinks]{hyperref}
\usepackage{gensymb}
\usepackage{amsmath}
\usepackage{dcolumn}    
\usepackage{ifpdf}
\usepackage{amssymb,lineno,amsfonts}
\usepackage{graphicx}   
\usepackage{bm}         
\usepackage{bbm}
\usepackage{mathrsfs}
\usepackage{upgreek}
\usepackage{mathtools}
\usepackage{epstopdf}
\usepackage{setspace}
\usepackage{hyperref}
\usepackage{natbib}
\usepackage{esvect}
\usepackage[usenames,dvipsnames]{xcolor}
\definecolor{med-blue}{RGB}{25,25,112}
\hypersetup{colorlinks, linkcolor={blue},citecolor={blue}, urlcolor={blue}}
\usepackage{times	}
\usepackage [english]{babel}
\usepackage [autostyle, english = american]{csquotes}
\MakeOuterQuote{"}

\begin{document}
\title{Quasi Static Remanence in  Dzyaloshinskii-Moriya Interaction driven Weak Ferromagnets and Piezomagnets}
\author{Namrata Pattanayak}
\affiliation{Department of Physics, Indian Institute of Science Education and Research, Dr. Homi Bhabha Road, Pune 411008, India}
\author{Arpan Bhattacharyya}
\affiliation{Saha Institute of Nuclear Physics, 1/AF Bidhannagar, Kolkata, India }
\author{A. K. Nigam}
\affiliation{Department of Condensed Matter Physics and Material Science, Tata Institute of Fundamental Research, Dr. Homi Bhabha Road, Mumbai 400 005, India}
\author{ Sang-Wook Cheong}
\affiliation{ Rutgers Center for Emergent Materials and Department of Physics and Astronomy, Rutgers University, Piscataway, New Jersey 08854, USA}
\author{Ashna Bajpai}
\affiliation{Department of Physics, Indian Institute of Science Education and Research, Dr. Homi Bhabha Road, Pune 411008, India}
\affiliation{Center for Energy Science, Indian Institute of Science Education and Research, Dr. Homi Bhabha Road, Pune 411008, India}

\date{\today}
\email[]{ashna@iiserpune.ac.in}

\begin{abstract}

We explore remanent magnetization ($\mu$) as a function of time and temperature, in a variety of rhombohedral antiferromagnets (AFM) which are also weak ferromagnets (WFM) and piezomagnets (PzM). These measurements, across samples with length scales ranging from nano to bulk, firmly establish the presence of a remanence that is quasi static in nature and exhibits a counter-intuitive magnetic field dependence. These observations unravel an ultra-slow magnetization relaxation phenomenon related to this quasi static remanence. This feature is also observed in a defect free single crystal of $\alpha$-Fe$_2$O$_3$, which is a canonical WFM and PzM. Notably, $\alpha$-Fe$_2$O$_3$ is not a typical geometrically frustrated AFM and in single crystal form, it is also devoid of any size or interface effects, which are the usual suspects for a slow magnetization relaxation phenomenon. The underlying pinning mechanism appears exclusive to those AFM which are either symmetry allowed WFM, driven by Dzyaloshinskii-Moriya Interaction (DMI) or can generate this trait by tuning of size and interface. The qualitative features of the quasi static remanence indicate that such WFM are potential piezomagnets, in which magnetization can be tuned by \textit{stress} alone. 

\end{abstract}

\pacs{Valid PACS appear here}
\maketitle
\section{Introduction}

\begin{figure*}[!t]
\includegraphics[width=1\textwidth]{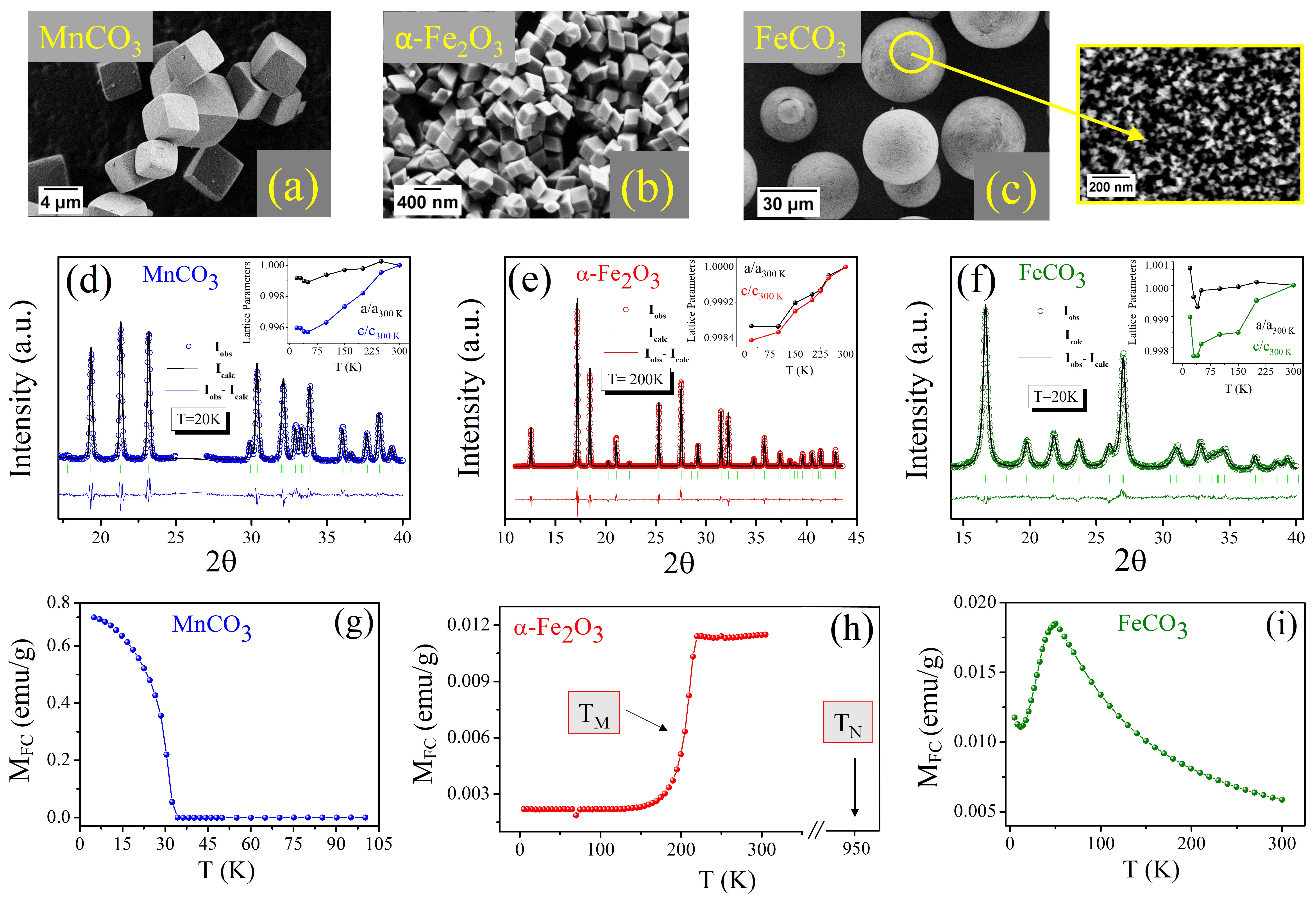}
\caption { SEM images of  \textbf{(a)} micro-cubes MnCO$_3$,  \textbf{(b)}nano-cubes $\alpha$-Fe$_2$O$_3$ and  \textbf{(c)}  FeCO$_3$ spheres with diameter $\thicksim$ 20 $\mu$m. Each sphere consists of triangular FeCO$_3$ nano particles $\thicksim$ 5-10 nm. Figures \textbf{(e)-(f)} display  Synchrotron  XRD data of MnCO$_3$ $\alpha$-Fe$_2$O$_3$ and FeCO$_3$  along with  Rietveld fitting. The inset shows best fit lattice parameters derived from Rietveld profile refinement of I vs 2$\theta$ data recorded at different temperatures.  The magnetization as a function of temperature  from 300 K to 5K in presence of $H$ = 1 kOe for all three samples is shown in figures \textbf{(g)}-\textbf{(i)}.  For MnCO$_3$ and FeCO$_3$ , the Neel transition temperature is 30 K and 50 K respectively as evident from \textbf{(g)} and  \textbf{(i)} respectively. For $\alpha$-Fe$_2$O$_3$, the Morin transition T$_M$ signifying spin re-orientation transition from  WFM to pure AFM state is around 260K. The actual Neel transition is around 950 K, shown schematically in  \textbf{(h)}.}
\label{Figure1}
\end{figure*}

Phenomenon of weak ferromagnetism in certain antiferromagnets, including the classic case of $\alpha$-Fe$_2$O$_3$, is associated with the experimental observation of a ferromagnetic (FM) like, spontaneous moment. This feature was initially attributed to a FM impurity phase in an otherwise AFM lattice; such as Fe$_3$O$_4$ impurity in $\alpha$-Fe$_2$O$_3$ \cite{Dzy1,Dzy2,Dzy3,Moriya1}. This controversy was firmly resolved by Dzyaloshiskii in 1958 \cite{Dzy1}, who proposed a spin canting mechanism that leads to a weak FM like state and Moriya \cite{Moriya1} who discovered the microscopic origin of this spin canting and its connection with spin orbit coupling (SOC). This is the celebrated  Dzyaloshinskii-Moriya Interaction (DMI), of the type  \textbf{D}.(\textbf{S}$_i$ X \textbf{S}$_j$) which is now central to both fundamental and application based trends in contemporary condensed matter physics. Apart from exotic inhomogeneous spin textures and non collinear spin systems such as skyrmions, topological insulators and superconductors, DMI/SOC also brings into fore the role of antiferromagnetic insulators in spintronics \cite{Rozler,Onose,Hasan,Yamasaki,Binz,Kane,Jairo, Barthem,Gross,Gayles}.

In many of the symmetry allowed weak ferromagnets, which include rhombohedral AFMs like $\alpha$-Fe$_2$O$_3$, MnCO$_3$ and rutile AFMs like NiF$_2$ or CoF$_2$, the phenomenon of stress induced moments or piezomagnetism, of the type (M$_i$ = P$_i$σ$_j$σ$_k$σ$\sigma$σ$_j$σ$_k$) where $\sigma$ is stress, was also predicted by Dzyaloshinskii \cite{Dzy1}. Experimental observations of such stress induced moments were made by Borovik-Romanov in a variety of WFM/PzM single crystals in the seminal work spanning from 1960s to 70s \cite{Romanov1,Romanov2,Romanov3,Romanov4}. On the similar lines of magnetoelectricity, wherein a magnetic moment can be created by \textit{electric field} alone - for which Cr$_2$O$_3$ is a prototype \cite{Birss, Halley}- magnetic moment from \textit{stress} alone can occur in PzM, for which $\alpha$-Fe$_2$O$_3$ is a prototype\cite{Birss,Sendonis,Paul}. It is also interesting that both Cr$_2$O$_3$ and $\alpha$-Fe$_2$O$_3$ are isostructural AFM but the piezomagnetic moments are observed in $\alpha$-Fe$_2$O$_3$, not in bulk Cr$_2$O$_3$. A picture also emerged with a plausible explanation on the microscopic mechanism of PzM in these systems\cite{Philip1,Philip2}.

In some of these WFM/PzM compounds or in their doped versions \cite{Kleemann}, an unusually slow magnetization relaxation was tracked through the measurement of remanence. This was further seen in ultra-thin films of Cr$_2$O$_3$ \cite{Binek}, in FM/AFM core shell systems where Cr$_2$O$_3$ appeared as an ultra-thin surface layer \cite{Ashna1} and also when Cr$_2$O$_3$ is encapsulated inside carbon nanotubes \cite{Ashna2} (CNT). These reports pointed towards some features in remanence which appear to be common, especially for AFM with the possibility of WFM/PzM. Most intriguing among these is ultra slow magnetization relaxation phenomenon, resulting in the observation of a quasi static remanence with a counter-intuitive  magnetic field dependence \cite{Ashna1,Ashna2}.

 \begin{figure*}[!t]
\includegraphics[width=1\textwidth]{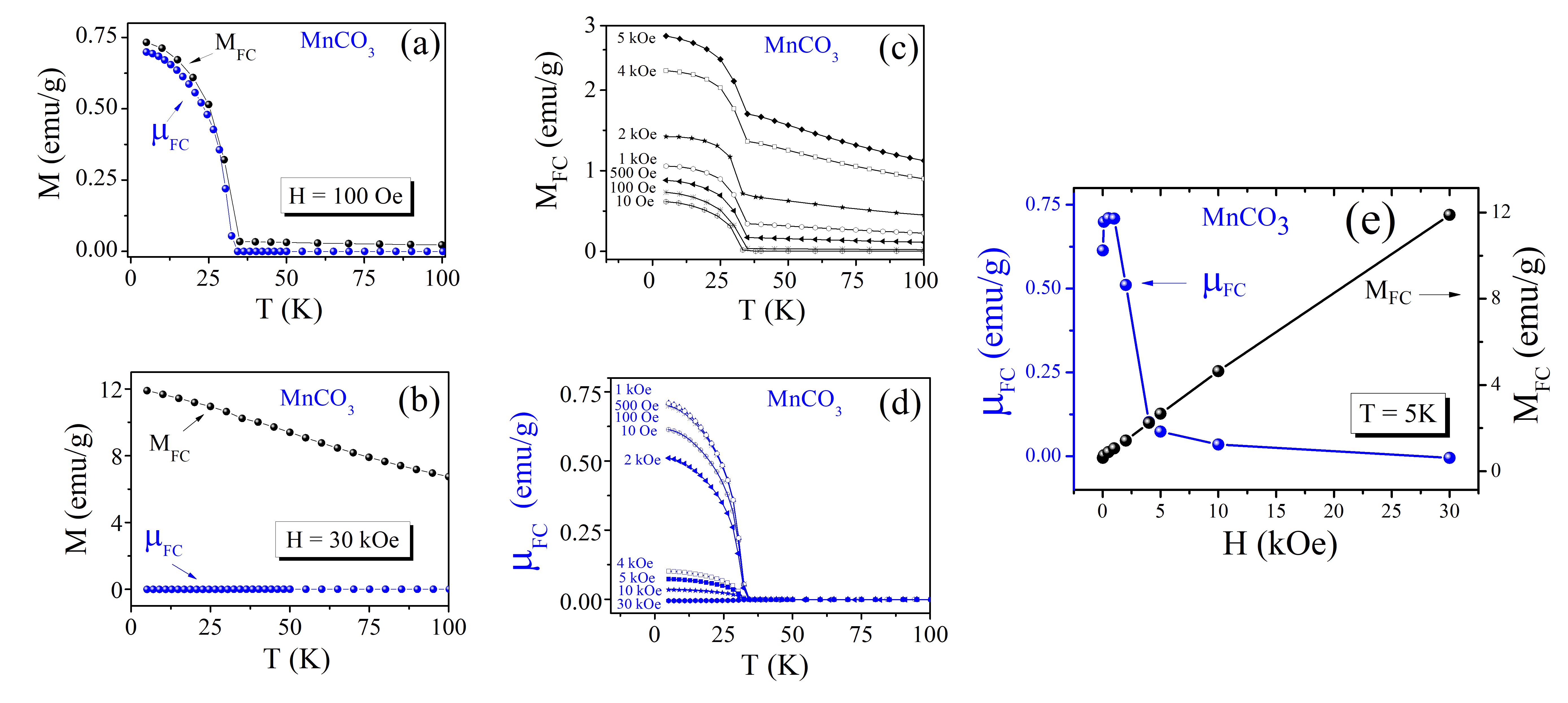}
\caption {Black dots in \textbf{(a)}  show magnetization measured while cooling ($H$ = 100 Oe) and  blue dots  are corresponding remanence ($H$=0) measured  while warming for MnCO$_3$ sample. \textbf{(b)} shows the same for $H$=30 kOe. \textbf{(c)} M$_{FC}$ vs T at various $H$ depicting regular AFM behavior with M$_{FC}$ rising with rise in $H$. \textbf{(d)} shows corresponding $\upmu_{FC}$ vs T exhibit s strikingly different cooling $H$ dependence. \textbf{(e)} compares the magnitude of  M$_{FC}$ (black dots, right axis) and $\upmu_{FC}$ (blue dots, left axis) at 5K as a function of (cooling) $H$ for MnCO$_3$.}
\label{Figure2}
\end{figure*}

Interestingly, Cr$_2$O$_3$ is not a symmetry allowed WFM/PzM but exhibits quasi static remanence only when it is in an ultra-thin form. It is therefore important to systematically explore whether these features intrinsically exist in symmetry allowed WFM and to investigate the circumstances in which this can appear in systems with altered symmetry conditions, especially due to size/interface effects.  In addition, what still remains an open question is whether piezomagnetism will always co-exist in all WFM and if so, what are the foot prints of this phenomenon? It is also important to explore possible means to isolate this subtle effect from routine magnetization measurements, wherein all other field dependent processes contribute for any AFM (canted or otherwise) under magnetic field. 

 In this work we explore remanence in two rhomohedral AFM that are symmetry allowed WFM and PzM. This includes $\alpha$-Fe$_2$O$_3$ with Neel transition temperature (T$_N$) $\thicksim$ 950 K and MnCO$_3$ with T$_N$σ $\thicksim$ 30 K.  Here $\alpha$-Fe$_2$O$_3$ is known to be a pure AFM upto 260 K and a WFM in the temperature range of 260-950 K \cite{Dzy1,Romanov2}. The temperature at which $\alpha$-Fe$_2$O$_3$ becomes WFM/PzM is also known as Morin Transition, T$_M$ ($\thicksim$ 260 K). It is advantageous to have a WFM near the room temperature for practical applications. However the effect is known to be much weaker than MnCO$_3$\cite{Dzy1}. We also investigate isostructural compound FeCO$_3$ with T$_N$σ $\thicksim$ 50 K, for which there are conflicting reports in literature about the existence of WFM and PzM \cite{Dzy1,Dzy2,Dzy3,Romanov3}.  For such cases, size effects may play a prominent role as DMI can be dominant and enhanced at surfaces and interfaces \cite{Mich}. 

 We study all three samples in the form of nano and mesoscopic crystals / particles and show a correlation between the structural parameters and the magnitude of pinned moment related to the quasi static remanence.  In case of $\alpha$-Fe$_2$O$_3$, which is also a prototypical PzM near the room temperature, we confirm the ultra slow magnetization relaxation in its single crystal form, thus bringing out that the quasi static remanence is intrinsic.  We also show that this feature can be substantially tuned by size effects, by  comparing the magnitude of quasi static remanence in the single crystal and nano cubes of  $\alpha$-Fe$_2$O$_3$. 

\section{Experimental Techniques}

Micro-cubes of MnCO$_3$ (length $\thicksim$ 2-4 $\mu$m), nano-cubes of  $\alpha$-Fe$_2$O$_3$ (length $\thicksim$ 200 nm) and polycrystalline spheres of FeCO$_3$ (grain size $\thicksim$ 5-10 nm) have been synthesized following the precipitation and hydrothermal routes \cite{Lis,Liu1,Liu2} Fig.\ref{Figure1}a-\ref{Figure1}c. The single crystal of $\alpha$-Fe$_2$O$_3$ has been grown using Floating Zone technique.  Scanning Electron Microscopy images are recorded using ZEISS ULTRA plus field-emission SEM. All the samples have been characterized using X-ray powder diffraction (XRD) using Bruker D8 Advance with Cu K$\alpha$ radiation ($\lambda$ = 1.54056 \AA). (\textbf{Supp. Info :} \textbf{Fig.~S1-S3}).  Temperature variation of synchrotron XRD from 20 K-300 K has been conducted in BL-18 beam line, Photon Factory, Japan. The synchrotron XRD data has been fitted using Rietveld Profile Refinement. All three samples stabilize in rhombohedral structure and fitting has been done in hex setting. The XRD data along with the Rietveld fittings at few selected temperatures for each of the sample is shown in Fig.\ref{Figure1}d-\ref{Figure1}f. The refined lattice parameters $a$ and $c$ at room temperature for all three samples  are given in table 1.  The Temperature variation of refined lattice parameters $a$ and $c$ for the samples are shown in the respective insets in Fig.\ref{Figure1}d-\ref{Figure1}f. Here both $a$ and $c$ are normalized with their respective room temperature value.  The magnetization measurements have been carried out by using a superconducting quantum interference device (SQUID) magnetometer, Quantum Design MPMS-XL.

\begin{table}

		\begin{tabular}{|c|c|c|c|}
		
		\hline
				Sample & $a$ (${\AA}$) & $c$ (${\AA}$) & $c/a$ \\
		\hline 
		 MnCO$_3$ & 4.7723(7) & 15.611(3) & 3.27\\
		
		\hline 
		FeCO$_3$ & 4.6678(4) & 15.202(1) & 3.25 \\
		
		\hline 
		$\alpha$-Fe$_2$O$_3$  & 5.0087(1) & 13.6856(4) & 2.73 \\
				\hline
				
	\end{tabular}
	\caption{Structural Parameters of MnCO$_3$,  $\alpha$-Fe$_2$O$_3$ and FeCO$_3$ as determined from the Rietveld analysis of room temperature X-ray diffraction data. }
	\label{Table1}
\end{table}

\section{Results and Discussions}

Magnetization  as a function of Temperature ( M$_{FC}$ vs T)  recorded  while cooling in presence of magnetic field $H$ = 1 kOe is presented Fig.\ref{Figure1}g-\ref{Figure1}i for all three samples. This is the routinely known Field Cooled (FC) cycle. The Neel transition temperature for MnCO$_3$ and FeCO$_3$  as shown in Figure 1, match well with the respective literature values. For both these samples, the $H$  is applied in the paramagnetic region, prior to the FC cycle. However, for  $\alpha$-Fe$_2$O$_3$, the T$_N$ is ~ 950 K and it is marked schematically in the \ref{Figure1}h. This is to indicate that in the case of $\alpha$-Fe$_2$O$_3$, the magnetization data is recorded while cooling from 300 K, which is above its Morin Transition temperature (T$_M$) but below its Neel temperature (T$_N$).  For single crystal of  $\alpha$-Fe$_2$O$_3$, the magnetization (M$_{ZFC}$) is also recorded in Zero Field Cooled (ZFC) state as would be shown in the latter part of the text. These factors have important implications while preparing a remanent state for all these samples considered here.

 \subsection{ Preparation of the Remanent State : FC/ZFC protocol }

Our primary tool here is DC magnetization in remanent state \cite{Morup,Ben1,Ben2,Neel,Binder,Mat,Suzuki}. This enables us to track the magnetization relaxation phenomenon and hence pinning potential landscape in all these WFM.  This remanent state is prepared   in two experimental protocols, the FC and ZFC.

In \textbf{FC} protocol, the sample is cooled in a specified magnetic field,  $H$, which  is applied  much above the  T$_N$ (or T$_M$) and the M$_{FC}$ is recorded while cooling.  The $H$ is switched off at 5K,  and  thereafter the remanent magnetization (or remanence) is experimentally measured in $H$ = 0 state. This  remanence , prepared after a typical FC cycle, is  referred to as \textbf{$\upmu_{FC}$}.  This can be measured either (i) as a function of increasing \textit{temperature}  from 5K to 300 K or (ii) as a function of \textit{time} at 5K.

In the \textbf{ZFC} protocol, employed only for the single crystal of  $\alpha$-Fe$_2$O$_3$,  the $H$ is applied from below the T$_M$ and M$_{ZFC}$ is measured in warming cycle , right upto 300 K. Thereafter $H$ is switched off and  the corresponding remanence, referred to as \textbf{$\upmu_{ZFC}$}, is measured as a function of \textit{time} at 300 K.

  We emphasize that in all the subsequent data involving \textbf{$\upmu$} presented in this work,  the magnitude of \textbf{$H$} indicated in the plots refers to the  magnetic field applied during  either cooling or warming cycle, so as to \textbf{prepare} a remanent state.  This remanence ($\upmu_{FC}$ or $\upmu_{ZFC}$ ), the origin of which is the subject matter of investigation here, is  experimentally measured  only after switching \textit{OFF} the $H$.

\subsection{Temperature Variation of Remanence in MnCO$_3$}

Fig.\ref{Figure2}a shows M$_{FC}$ vs T, (measured while cooling)  in presence of $H$ $\thicksim$ 100 Oe (black dots). The magnitude of M$_{FC}$  at 5K $\thicksim$ 0.75 emu/g. After removal of $H$ at 5K, a part of the magnetization decays instantaneously. However, a significant part of magnetization remains pinned, resulting in the observation of remanence.  This remanence ($\upmu_{FC}$) shows almost no further decay as a function of time, as long as the temperature is held constant at 5K. As evident from Fig.\ref{Figure2}a, the magnitude of the $\upmu_{FC}$ at 5K is  $\thicksim$ 0.7 emu/g for this run.  On increasing the temperature, $\upmu_{FC}$ vs T (measured while warming) shows a variation which is qualitatively similar to M$_{FC}$ vs T right up to the T$_N$  as shown in Fig.\ref{Figure2}a (blue dots). In the paramagnetic tail, the $\upmu_{FC}$ vanishes, as is expected. 

Fig.\ref{Figure2}b shows the same for $H$ $\thicksim$ 30 kOe, for which M$_{FC}$ $\thicksim$ 12 emu/g whereas the $\upmu_{FC}$ $\thicksim$ $10^{-5}$ emu/g at 5 K. Thus the $\upmu_{FC}$ is vanishingly small for 30 kOe run. We would consider the $\upmu$ of this magnitude to be roughly arising from the quenched field of SQUID superconducting magnet, which may be $\thicksim$ 5-10 Oe and can vary from run to run\footnote{In SQUID magnetometers, even in $H$ = 0, there  can be some residual magnetic field arising from the superconducting coil. The magnitude of this residual field can be 5 to 10 Oe and its sign can be arbitrary. The vanishingly small value of remanence at $H$ = 30 kOe  also sets the base line for any artifacts arising from such residual fields.}. The data contained in Fig.\ref{Figure2} clearly indicates that the magnitude of $\upmu$ is almost equivalent to that of M$_{FC}$ for lower (cooling) $H$ whereas it is negligible for very high $H$. 

\begin{figure}[!t]
\hspace*{-1.0cm}
\includegraphics[width=0.5\textwidth]{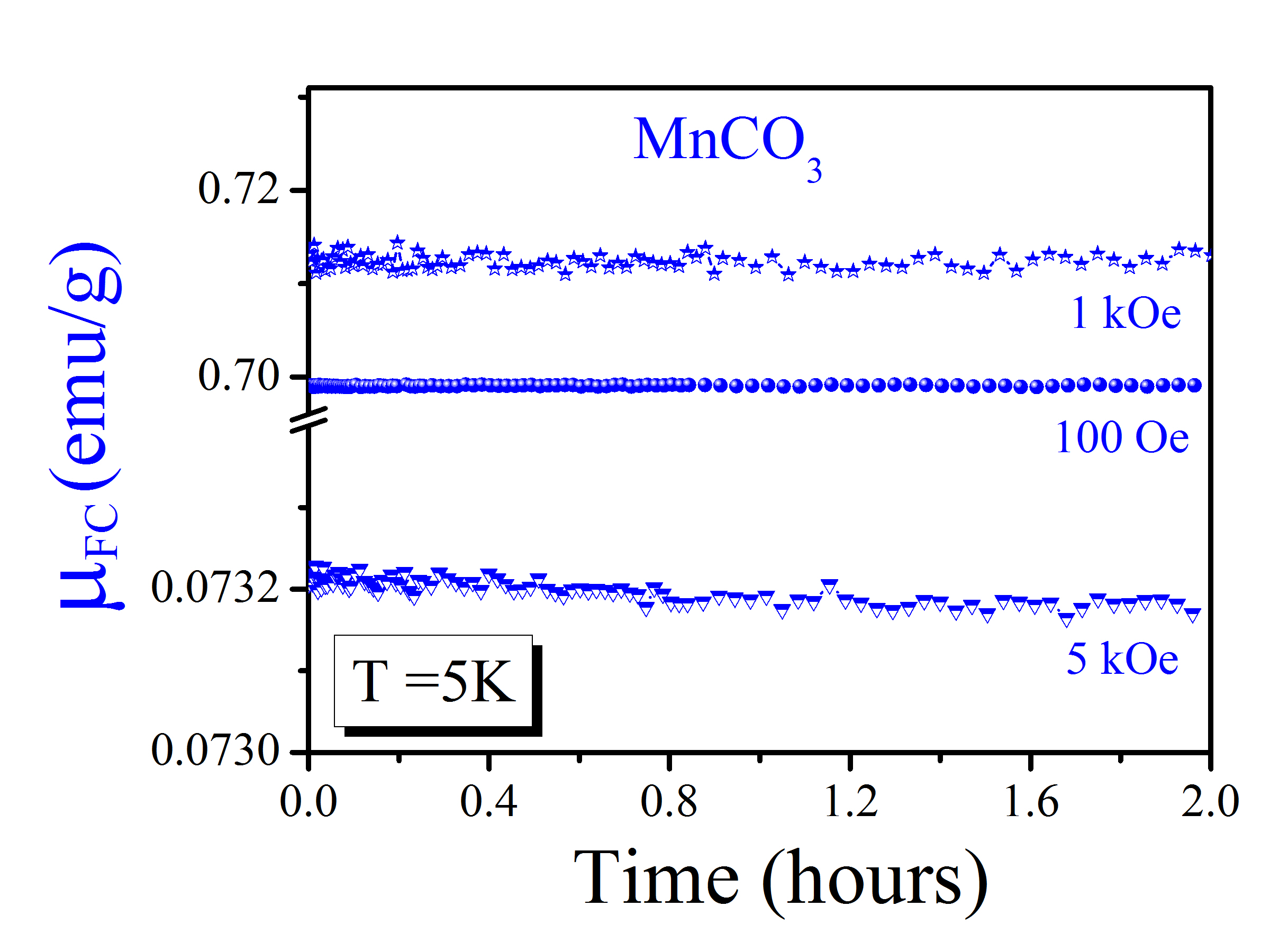}
\caption {Remanence as a function of time for three different cooling fields at a fixed temperature of 5K for MnCO$_3$. These data show that the remanence is almost constant  over a time period of 2 hours, thus  depicting its  quasi static nature.}
\label{Figure3}
\end{figure}

For all the intermediate magnetic fields, the M$_{FC}$ vs T data are plotted in \ref{Figure2}c and their corresponding $\upmu_{FC}$ vs T are plotted in Fig.\ref{Figure2}d. As is evident from these data, the magnetization  increases with increasing $H$, consistent with a regular AFM behaviour.  However, the corresponding remanence varies with the strength of the magnetic field in an unexpected way. Here the remanence is first seen to rise with increasing $H$, upto a critical field. Thereafter it decreases with increase in field and eventually vanishes beyond another critical field.

\begin{figure*}
\includegraphics[width=1\textwidth]{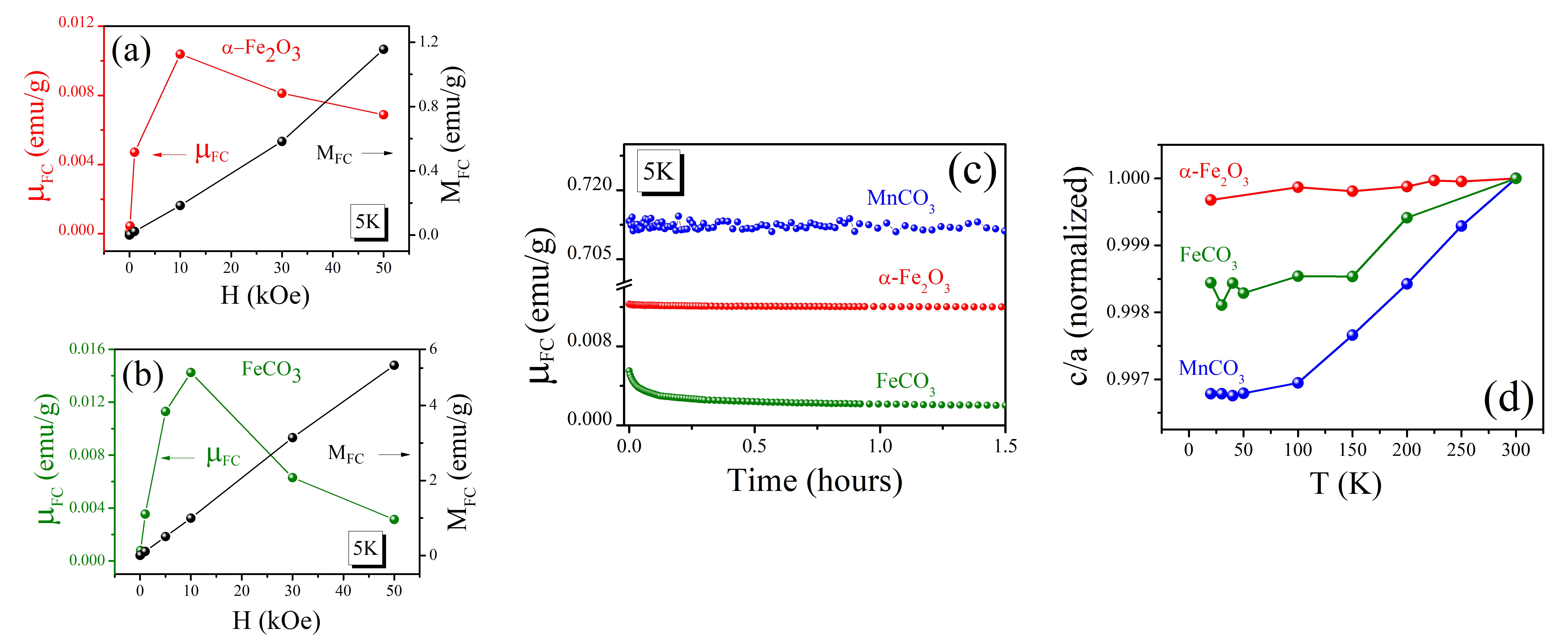}
\caption{\textbf{(a)} and \textbf{(b)}  shows  $\upmu_{FC}$  as a function of (cooling) $H$ for $\alpha$-Fe$_2$O$_3$ and FeCO$_3$ respectively. The corresponding  M vs H is shown for each sample is shown in the same graph, indicating the unusual (cooling) field dependence of remanence for both the samples.   \textbf{(c)}  Compares $\upmu_{FC}$ as a function of $time$  in all three samples. The observation of the  quasi static nature of remanence is unambiguous in case of MnCO$_3$  as well as $\alpha$-Fe$_2$O$_3$. \textbf{(d)} shows $c/a$ ratio using refined lattice parameters obtained at various temperature from Rietveld fitting of synchrotron XRD data.  The $c/a$ ratio has been normalized with its value at 300 K.}
\label{Figure4}
\end{figure*}

To clearly bring out the unusual (cooling) field dependence of the $\upmu_{FC}$, we compare the magnitude of both M and $\upmu$  at 5K. These data points are extracted from different M$_{FC}$ vs T  and their corresponding $\upmu_{FC}$ vs T runs Fig.\ref{Figure2}e. Here M$_{FC}$ is seen to increases with increasing $H$, as is expected for a regular AFM, whereas the $\upmu_{FC}$ initially rises with increasing $H$, followed by a sharp drop. The $\upmu_{FC}$ completely vanishes at very high field.  The type of field dependence of  $\upmu$ is not expected for either a regular FM or AFM\cite{Ben1,Ben2,Neel}. Thus the $H$ dependence of the remanence (blue dots)  brings forward a unique functional form, which is not observed in the routine M vs $H$ isotherm (black dots) .

\subsection{Remanence in MnCO$_3$: Variation with Time}

 To check the stability of the remanence as a function of time, we also performed relaxation rate measurements. After a typical M$_{FC}$ vs T and subsequent removal of $H$, we  obtained $\upmu_{FC}$ vs \textit{time}, while the temperature is held constant at 5K (Fig.\ref{Figure3}).These remanence data, obtained for three different cooling fields, again brings forward two distinct magnetization relaxation rate, one of which is ultra-slow.  We observe that for measurement times of about two hours, the $\upmu_{FC}$ shows no appreciable decay and this type of  remanence can be termed as quasi static in nature. 

Consistent with the data presented in Fig.\ref{Figure2}d, magnitude of the $\upmu_{FC}$ is seen to vary with cooling field $H$ in a way, which is not obvious from the routine temperature M-H isotherms. For the chosen cooling fields of 100 Oe, 1 kOe and 5 kOe, the $\upmu_{FC}$  values are $\thicksim$ 93\%, 70\% and 3\% of their corresponding M$_{FC}$ values. These data also indicate that finding an optimum value of the (cooling) magnetic field enables almost all the in-field magnetization to be retained. For instance,  the remanence corresponding to 100 Oe run is ~ 93\%  of its M$_{FC}$ value. However, even for the run corresponding to 5 kOe, for which the magnitude of remanence is about  3\% of its  M$_{FC}$ value, the relaxation rate is still ultra slow.  Thus the data contained in Fig.\ref{Figure3} confirms presence of the remanence that is quasi static in nature with ultra-slow magnetization dynamics, and exhibits a counter intuitive $H$ dependence (Fig.\ref{Figure2}e).

\subsection{Remanence and structural parameters in $\alpha$-Fe$_2$O$_3$,FeCO$_3$ and MnCO$_3$}

Similar measurements were also conducted for $\alpha$-Fe$_2$O$_3$ and FeCO$_3$ samples.  Fig.\ref{Figure4}a and Fig.\ref{Figure4}b displays $\upmu_{FC}$ vs H data at 5K (extracted from various $\upmu_{FC}$ vs T runs) for both the samples. These data  reveal that the $\upmu_{FC}$ vs  $H$ for each of the sample is strikingly different from  corresponding M$_{FC}$ vs $H$ shown on the right axis in each plot.  Both the samples  exhibit a sharp rise in $\upmu_{FC}$ as a function of (cooling) $H$ and the peak value of $\upmu$ is obtained at different critical $H$ for each sample. This rise is qualitatively similar to  what is seen for MnCO$_3$ (Fig.\ref{Figure2}e), though  the fall, after the peak is  not as rapid.  Overall, the field dependence of remanence is counter-intuitive in all three samples.    

 In addition, all three samples exhibit two distinct time scales for magnetization decay, one of which is ultra slow and can be termed as quasi static. This slow magnetization  relaxation is evident in $\upmu_{FC}$ vs time measurements as shown in Fig.\ref{Figure4}c.   For  sake of comparison,  for each sample the remanent state is prepared in cooling magnetic field of 1 kOe.  The magnitude of the remanence is atleast an order of magnitude higher for MnCO$_3$ as compared to $\alpha$-Fe$_2$O$_3$.  This is also consistent with the earlier observations which indicate that the net FM moment  due to spin canting is about an order of magnitude larger larger in MnCO$_3$ \cite{Dzy1}. 

  To correlate the observed features in $\upmu$ with structural parameters, the temperature variation of $a$ and $c$ lattice parameters is studied.  As can be seen from the inset of Fig.\ref{Figure1}d, for MnCO$_3$, both $a$ and $c$ decrease with reducing temperature monotonically till about the T$_N$, however an expansion in both the lattice parameters is observed just below its AFM transition temperature. In addition, for MnCO$_3$ the lattice parameter $c$ is seen to fall much rapidly with reducing temperature as compared to $a$ (inset of Fig.\ref{Figure1}d). On the contrary, for $\alpha$-Fe$_2$O$_3$, the pattern of temperature variation for $c$ and $a$ are quite similar in nature and a slight trend of expansion in both lattice parameters is observed around its WFM region (Fig.\ref{Figure1}e). For all three samples, both lattice parameters exhibit a slight anomaly below T$_N$ (or around WFM in case of $\alpha$-Fe$_2$O$_3$), however the effect is more pronounced for the MnCO$_3$. 
	
	Fig.\ref{Figure4}d compares normalized $c/a$ ratio as a function of temperature for all three samples. This normalization is w.r.t $c/a$ ratio at 300 K for each sample. We find that the $c/a$ ratio shows a more rapid decline with reducing temperature  and a clear anomaly  is observed in the WFM region for MnCO$_3$. This trend also correlates with the stability and magnitude of the $\upmu$, both of which are relatively higher for MnCO$_3$ as compared to $\alpha$-Fe$_2$O$_3$. In case of FeCO$_3$, though the qualitative features in remanence  are similar,  but  the morphology of the sample makes its difficult to conclude whether these features are intrinsic or arising due to nano scaling.  In this case, the grain size is of the order of 2-5 nm,  Fig.\ref{Figure1}c. This situation is similar to what is observed for Cr$_2$O$_3$ which is also isostructural with $\alpha$-Fe$_2$O$_3$ but it is not symmetry allowed WFM in bulk. However, it exhibits slow relaxation and the unusual cooling field dependence of remanence only in ultra thin form \cite{Ashna1,Ashna2}.  Microscopic measurements are needed to confirm the presence of WFM in such cases, including ultra small FeCO$_3$ grains used in this study.    

 \begin{figure*}[!t]
\includegraphics[width=1\textwidth]{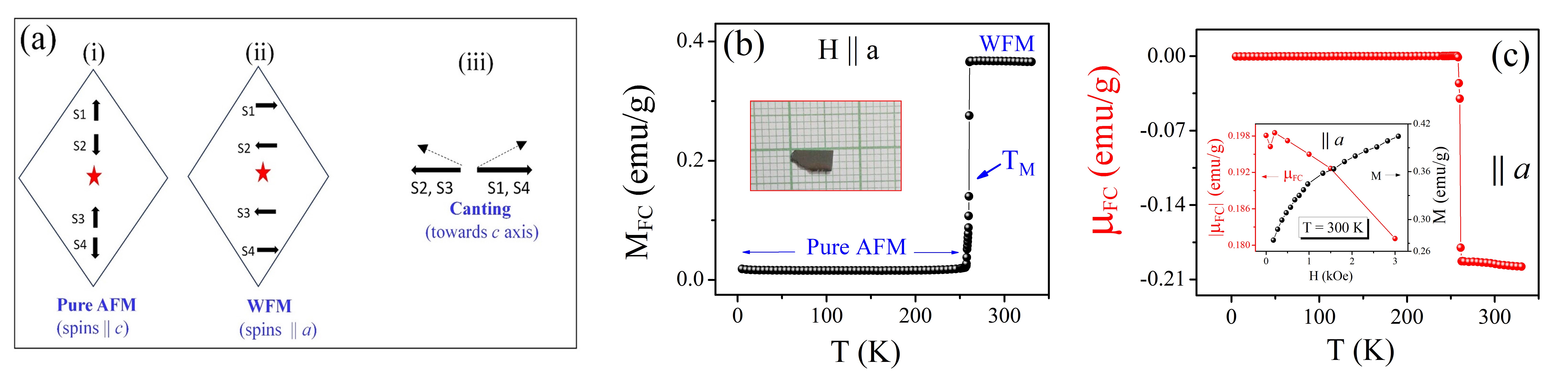}
\caption{ \textbf{(a)} shows schematic of typical  spin configurations in pure AFM (i) and WFM (ii) phase along with the phenomenon of canting (iii). The red star in (i) is the inversion center and the spins point along the $c$ axis in pure AFM phase.  In WFM phase, spin tilt in basal plane as shown in (ii). The spin configuration shown in (ii) is necessary for the observation of DMI driven spin canting that results in WFM phase. \textbf{(b)} M$_{FC}$ vs T  for a single crystal of $\alpha$-Fe$_2$O$_3$, with $H$ (1kOe)  parallel to $a$ axis, exhibiting  $T_M$,  the  Morin transition, marked as blue arrow in the figure.  The inset shows the picture of the $\alpha$-Fe$_2$O$_3$ single crystal. \textbf{(c)} shows $\upmu_{FC}$ vs $T$  run corresponding to the M$_{FC}$ vs T run shown in \textbf{(b)}.  Here the remanence is vanishingly small in the pure AFM region and finite in the WFM region. The inset shows $\upmu_{FC}$  at 300 K  along $a$ axis as a function of various (cooling) $H$. These data points are extracted from various $\upmu_{FC}$ vs $T$ runs. }
\label{Figure5}
\end{figure*}

For physical mechanism related to the remanence  that results in ultra-slow magnetization relaxation, a number of phenomena such as glassy phase, superparamagnetism, defect pinning in a regular FM or AFM, exchange bias at FM/AFM interface etc. can be considered. Such phenomena are known to result in slow relaxation with a variety of temporal functional forms\cite{Morup,Ben1,Ben2,Neel,Binder,Mat}.  However the mechanism behind the  quasi static remanence and its unusual  (cooling) magnetic field dependence  in these samples appears to be different from above mentioned phenomena. For instance, considering size effects,  MnCO$_3$ shows most robust magnetization pinning at  lower fields, as shown in Fig.\ref{Figure3}. However, the sample used for magnetization measurements consists of fairly big crystallites ($\thicksim$ 2-4 $\mu$m) therefore it is less likely that the slow relaxation is arising from size reduction or nano scaling. It is neither a glassy system, nor a nano scale FM which can exhibit superparamagnetic traits. Crystallites are also regular shaped with well-formed facets therefore the phenomenon of  defect pinning  leading to  ultra-slow magnetization relaxation is ruled out.  Also, for a regular AFM/FM, the $\upmu$ should have shown saturation \cite{Ben1} with $H$, rather than the sharp drop such as seen in Fig.\ref{Figure2}e.

To understand the nature of remanence in AFM with  WFM traits and to confirm if this effect is intrinsic, we also explored it in a single crystal (SC). For this purpose, we chose a SC of $\alpha$-Fe$_2$O$_3$ as this sample is well known to exhibit a spin reorientation transition from pure AFM to WFM phase\cite{Dzy1}.   

\subsection{ Pure AFM and WFM Phase  : Symmetry Considerations }

 Among the samples considered here, $\alpha$-Fe$_2$O$_3$ is known to be both pure AFM (upto 260 K) and WFM (260 K - 950 K) \cite{Dzy1}.  Here pure AFM phase implies that the DMI driven spin canting is not symmetry allowed. As mentioned before, isostructural compound Cr$_2$O$_3$  which does not exhibit spin canting,  in this context, is a pure AFM phase\cite{Dzy1}. For the sake of clarity, the spin configurations in  pure-AFM and WFM state are compared  in Fig.\ref{Figure5}a. In pure AFM phase, the spins within unit cell are arranged  along $c$ axis as shown in Fig.\ref{Figure5}a, configuration (i). Here the red star is the  inversion center and the spin configuration can be S1 = -S2 = S3 = -S4 as shown in (i). In WFM state, the spins re-orient to the basal plane, arranged in a specific sequence, in which S1=-S2=-S3=S4.   It is important to note that the unit cell is still AFM, but the spin configuration shown in (ii) is essential for DMI driven spin canting.   This  \textbf{D}.(\textbf{S}$_i$ X \textbf{S}$_j$) type of interaction  is possible between  sub-lattices associated with antiferromagnetically coupled spins, with the sign of D consistent with the symmetry considerations discussed in references 1-4.  The direction of the net FM moment due to the  spin canting  is towards the $c$ direction as is shown schematically  in Fig.\ref{Figure5}a(iii). This net FM moment in \textbf{otherwise AFM }is  responsible for \textbf{weak ferromagnetism}.   

The  spin configurations shown in Fig.\ref{Figure5}a(ii) is valid for all the rhombohedral AFM discussed here,  which are symmetry allowed WFM. For $\alpha$-Fe$_2$O$_3$,  the spin reorientation transition from pure AFM (spins along $c$ axis)  to WFM state (spins along $a$ axis)  occurs at T$_M$, the Morin transition temperature\cite{Dzy1}. Thus $\alpha$-Fe$_2$O$_3$ provides a unique opportunity to probe both AFM and WFM phase in the same sample, which individually exist in a wide temperature range. In the following, we present results of remanence measurements in the single crystal of $\alpha$-Fe$_2$O$_3$ in both the regions.

\subsection{Remanence in a Single Crystal of $\alpha$-Fe$_2$O$_3$ : Variation with Temperature}

Main panel of Fig.\ref{Figure5}b shows M$_{FC}$ Vs T for a SC of $\alpha$-Fe$_2$O$_3$ sample along $a$ axis. The Morin transition at $\thicksim$ 260 K demarcates the two regions, pure AFM and WFM for this sample. From  this, we note that magnitude of M $_{FC}$ is roughly $\thicksim$ 0.35 emu/g in WFM region and $\thicksim$ 0.015 emu/g in the pure AFM region.  After switching off the field at 5K, corresponding $\upmu_{FC}$ vs T in warming cycle is shown in the main panel of Fig.\ref{Figure5}c.  The $\upmu_{FC}$ is found to be negligibly small ($10^{-5}$ emu/g) in the pure AFM region and substantially large in WFM region (-0.2 emu/g). 

 Here, the sign of the $\upmu_{FC}$ is found to be negative w.r.t the direction of applied $H$.  From a number of such $\upmu_{FC}$ vs T data along $a$ axis, we find that the sign of $\upmu_{FC}$ at 300 K remains primarily negative  and its magnitude shows a slight decrease with increasing magnetic fields (inset in Fig.\ref{Figure5}c). It is to be noted that  for obtaining this data, the $H$ during FC cycle is applied at 300 K, when the sample is in WFM region. This is unlike the case of MnCO$_3$, where the $H$ can be applied in the paramagnetic region. For obtaining the (cooling) field dependence of remanence unambiguously, such as shown in Fig.\ref{Figure2}e for MnCO$_3$, it is preferable to apply the $H$ in the paramagnetic region for preparing individual remanent states.  However, in case of $\alpha$-Fe$_2$O$_3$, it is not practically possible to heat the sample above 950K, after each run. Though the sign of the $\upmu_{FC}$ along $a$ axis is not commensurate with the direction applied $H$ while cooling, its magnitude is substantial only in WFM region. 

To check the stability of this remanence as a function of \textit{time}, we conducted relaxation measurements both along the $c$ as well as $a$ axis.  Since the direction of net FM moment is likely to be towards the $c$ axis of the crystal, we particularly checked  stability of  $\upmu_{ZFC}$ as well as  $\upmu_{FC}$ along $c$ axis as a function of \textit{time}.

\begin{figure}
   \vspace*{0.6cm}
	\hspace*{-0.1cm}
	\includegraphics[scale=0.35]{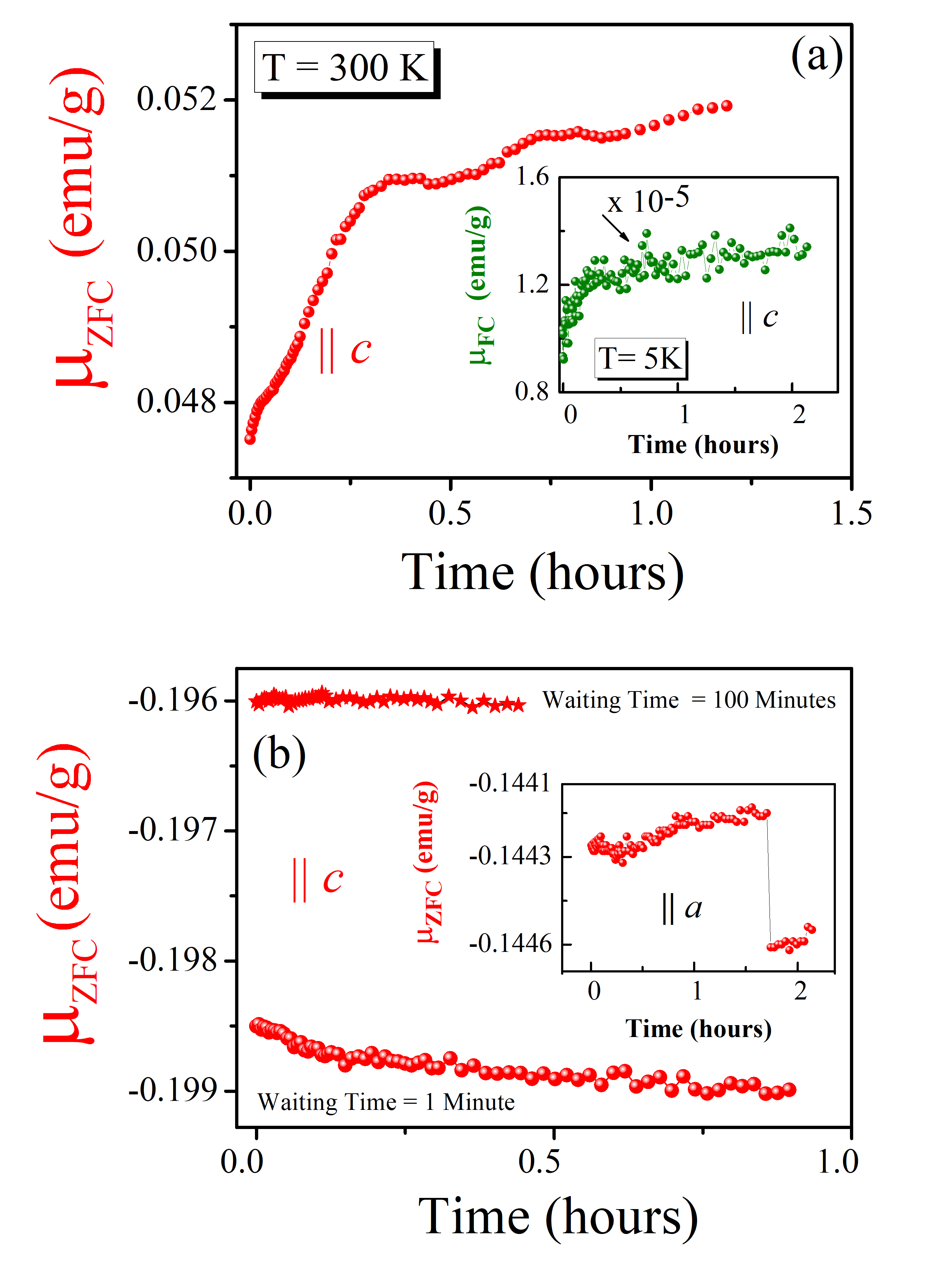}
	\caption{ \textbf{(a)} shows $\upmu_{ZFC}$ vs time at 300 K (WFM region) measured along the $c$ axis for the single crystal of $\alpha$-Fe$_2$O$_3$ . The inset shows $\upmu_{FC}$vs time along $c$ axis at 5K (pure AFM region). While  $\upmu_{FC}$ is negligibly small in the pure AFM region, it is substantially large (at-least by a few orders of magnitude) in the WFM region. \textbf{(b)} Main panel shows $\upmu_{ZFC}$ vs time measurements  parallel to $c$ axis for waiting time of 100 minutes (red stars) and 1 minute (red dots).  Inset shows $\upmu_{ZFC}$ vs time parallel to $a$ axis showing discrete jump.}
\label{Figure6}
\end{figure}

\subsection{Remanence in a Single Crystal of $\alpha$-Fe$_2$O$_3$ : Variation with Time}

In this section we present relaxation rate of remanence in pure AFM and WFM phase of $\alpha$-Fe$_2$O$_3$,  measured following the FC and ZFC protocol respectively. 

For   remanence in the pure AFM region, the  $H$ is applied from 300 K and M$_{FC}$ vs T is recorded while cooling (not shown here). The $H$ is switched off at 5K and the $\upmu_{FC}$ is measured as a function of time. This is  shown in the inset of Fig.\ref{Figure6}a.  These data further confirm that the remanence is negligible  in pure AFM region $\thicksim$ $10^{-5}$ emu/g (inset of Fig.\ref{Figure6}a).

For preparing the remanent state in WFM region,  the  $H$ is applied from below the T$_M$ and M$_{ZFC}$ vs T  is recorded while warming , right upto 300 K (not shown here). At 300 K, the $H$ is switched off and $\upmu_{ZFC}$ is measured as a function of time (main panel, Fig.\ref{Figure6}a).  Here, the remanence is positive and is commensurate with the direction of $H$ applied during ZFC cycle. Thus the remanence is substantial in magnitude in WFM region and  it is also fairly stable in time. 

 However, from a number of $\upmu_{ZFC}$ vs time cycles in positive $H$, we observe that the magnitude of $\upmu_{ZFC}$ in WFM region varies from 0.05-0.2 emu/g but its sign primarily remains negative. This anomaly appears only in the remanence measurements but not in the regular in-field measurements such as shown in Fig.\ref{Figure5}b. However, such ambiguity with sign has also been observed in the sign of stress induced moments in some WFM/PzM on repeated cooling \cite{Romanov1}. The reason for such ambiguity in case of remanence, (which does not appear in regular \textit{in-field} magnetization) is also discussed in the latter part of the text.  We also note a slight variation (~5\%) in the magnitude of $\upmu$, from run to run, for the same  (cooling) magnetic field. These anomalies are also seen to appear only in the WFM region.

 Interestingly, we  also observe a slight trend of rise ( $\thicksim$ a few\% of total remanence) in $\upmu_{ZFC}$ vs time data,  as shown in Fig.\ref{Figure6}a.  The over all relaxation data appears to be a sum of both  time-decay as well as time-rise of the remanence. This  indicates that on application of $H$ (while preparing the remanent state) the moments continue to reorient slowly in presence of $H$ and on the removal of $H$, the time decay is ultra slow as well. This also indicates that the total time span  in which the $H$ is ON for  preparing a particular  remanent state is also an important parameter.  This could  also be responsible for variations in the magnitude of the remanence, as observed here.  This result prompted us to perform \textit{waiting time} dependence, usually employed for glassy systems\cite{Binder}.  

For waiting time runs,  two  remanent states are prepared  using the same (cooling) magnetic field. In first case, the $H$ =1 kOe is applied  in ZFC protocol, from below the  T$_M$ and the sample is heated right up to 300 K. At 300 K the magnetic field was kept ON for waiting-time of 1 minute, prior to finally switching it OFF for the remanence measurements. The second remanent state is prepared following exactly the same protocol, however this time the  $H$ = 1 kOe  is kept ON for waiting time of 100 minutes, prior to switching it OFF.  These $\upmu_{ZFC}$ vs $time$  data parallel to $c$ axis  are presented  in the main panel of Fig.\ref{Figure6}b,  for 1 min (dots) or 100 min (stars) waiting-time respectively. These data  clearly indicate that the magnitude of the remanence  also  changes with the total time span of the $H$ applied for preparing a particular remanent state. This also  explains the the slight differences in the magnitude of remanence from run to run. The inset shows the same for $\upmu_{ZFC}$ parallel to $a$ axis after 100 min of waiting time. Along the $a$ axis, the magnetization  relaxation is ultra slow and occasionally discrete jumps in  remanence are observed, though the change is less than a percent. However, the remanence  continues to  exhibit quasi static  nature.  

 These anomalies  which exist in the remanent state  are not observed in routine M vs T measurements.  $\alpha$-Fe$_2$O$_3$ is not a frustrated AFM  and in the single crystal form, size/interface related phenomena cannot account for the \textit{waiting time} effects and ultra-slow magnetization dynamics. From the observation of quasi static remanence in single crystal, together with similar features observed in MnCO$_3$, we conclude that the ultra-slow magnetization dynamics can be taken as indicative of the presence of WFM. This ultra-slow dynamics also appears to be associated with the microscopic details of the AFM domain which turn WFM due to spin canting.

\subsection{Quasi static Remanence and DMI driven Spin Canting }

  Considering the microscopic reason for  quasi static remanence  (that leads to the ultra-slow magnetization dynamics as observed here) in these systems, we recall the details of magnetic structure in all these compounds. The spin arrangement shown in Fig.\ref{Figure5}a(ii) is essential for the observation of WFM. This should also limit the possible ways in which an AFM domain can exist in the WFM region. For a regular AFM, on the application of the $H$, the induced magnetization is driven by the Zeeman energy and the magnetocrystalline anisotropy. However, the additional factor in WFM will include response from \textit{spontaneously} canted spins, related to the DMI as well. On removal of $H$, the reversal of the WFM domain will have to be accompanied by the reversal of the AFM moment which is energetically unfavorable \cite{Romanov1}. Once a AFM domain with  spin-canting is formed, guided by a cooling $H$ applied  from above the AFM to PM transition, it is energetically unfavorable for these domains to relax, when the $H$ is removed. This feature is only observed upto a critical value of $H$ which can vary depending on the sample, as is observed here (Fig.\ref{Figure3} and Fig.\ref{Figure4}a ). Beyond a critical $H$, the magnetization dynamics is driven by Zeeman and magnetocrystalline anisotropy. The magnetization relaxation in this case is much faster, similar to what is observed for a normal AFM.  However, below this critical field strength, the WFM domain configuration is guided by the sign of $H$ field, when, it is applied from T $>>$ T$_N$.   When the $H$ is applied in WFM region, the spins are already spontaneously canted.  This also explains the ambiguity with sign, as observed in case of $\alpha$-Fe$_2$O$_3$. 
 
\begin{figure}[!h]
\includegraphics[width=0.48\textwidth]{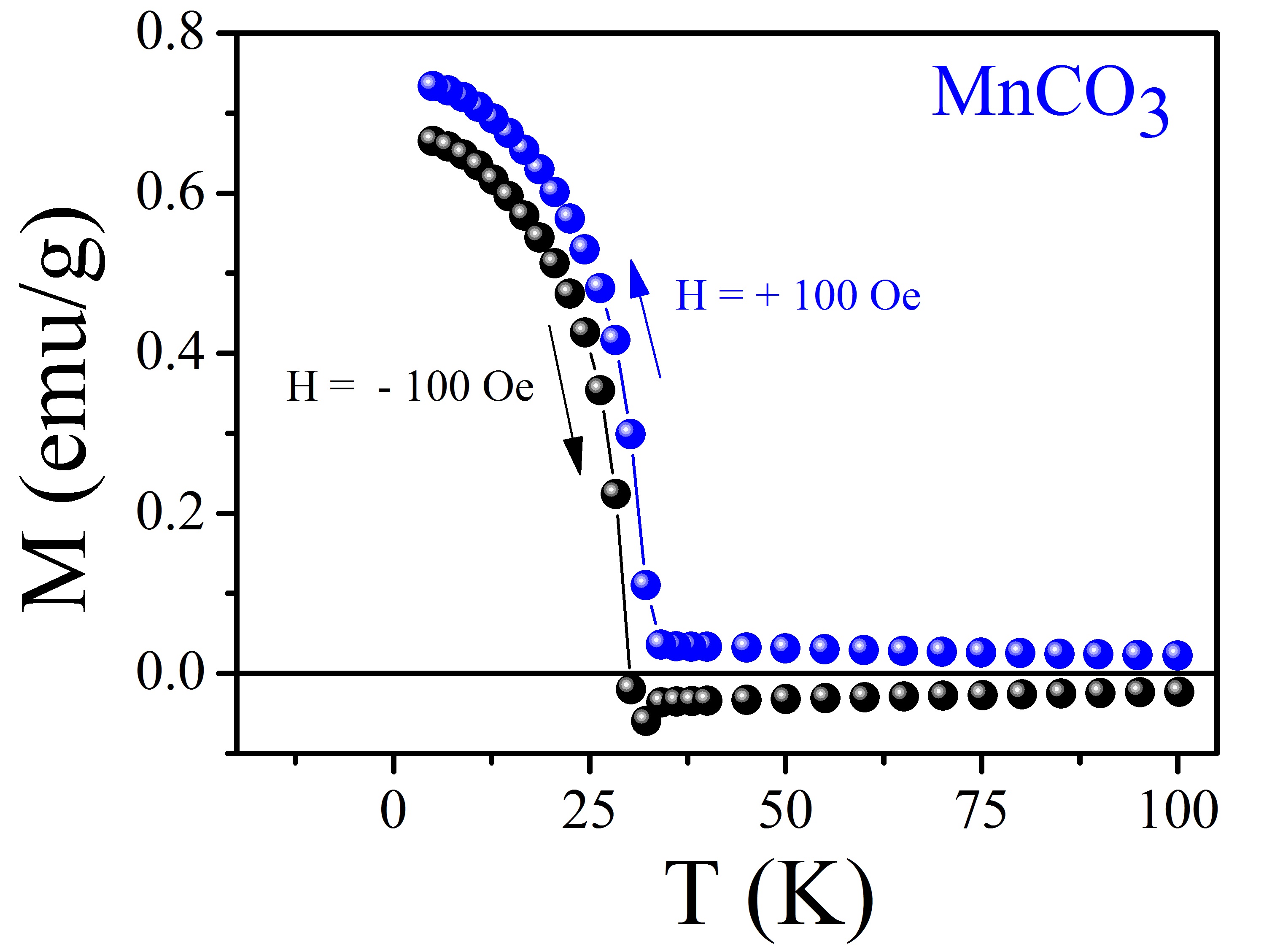}
\caption{M Vs T recorded while cooling in presence of H = + 100 Oe (blue dots). At 5K, the H = +100 Oe is removed and H = -100 Oe is applied while the temperature is held constant at 5K. Sunsequently M vs T in presence of H = -100 Oe is again recorded in warming cycle (black dots).  The measured magnetization is basically the remanence prepared during the previous cooling cycle. Since the sample is already in the WFM state, the presence of H = -100 Oe is not sufficient to rotate the magnetization. The robustness of pinned moment, which leads to quasi static remanence, for MnCO$_3$ is evident from this data.}
\label{Figure7}
\end{figure}

For further confirming that the ambiguity with sign is related to spontaneous spin canting related with DMI and not arising due to measurement related artifacts,  we  revert back to MnCO$_3$ which has a T$_N$ $\thicksim$ 30 K and $H$ can be applied in the paramagnetic region. Fig.\ref{Figure7}  shows M$_{FC}$ vs T data recorded while cooling from above T$_N$, down to 5 K,  in presence of $H$ = + 100 Oe (blue dots) . At 5K the $H$  is switched off and the quasi static remanence is observed, which is positive in magnitude as the WFM domain configuration is already guided by the $H$ = + 100 Oe.  Temperature still held at 5K, we  again apply  $H$ = -100 Oe and subsequent to this, the M vs T is measured in warming cycle (FH cycle) in presence of $H$ = -100 Oe.  As is evident from the data shown in Fig.\ref{Figure7}, once pinned in WFM state from above T$_N$ by a positive $H$, the negative field cannot change the sign of pinned moment and therefore the sign of remanence. The measured magnetization in presence of $H$ = -100 Oe while warming(black dots) is still positive and clearly a magnetic field applied in WFM region does not make any difference.  Thus the observed magnetization  is basically due to the presence of positive remanence, stabilized during  previous ($H$ = + 100 Oe) FC cycle.  This data explains the ambiguity related with the sign of remanence, especially when the $H$ field is applied in WFM region.

Over all, these data confirm that  the quasi static remanence  is observed below a critical value of $H$ in WFM and related to anisotropic exchange. At higher $H$, the interplay is between Zeeman and exchange energy, as is usually observed for a regular AFM. The ambiguity related with the sign of $\upmu$ in single crystal of $\alpha$-Fe$_2$O$_3$ is related with configuration of AFM domains  in which the spins are spontaneously canted due to DMI, even in the absence of $H$. On cooling or heating in presence of $H$ leads to stabilization of these canted AFM domains in different configurations, compatible with the interplay of various energy scales involved. This feature again indicates that the net moment related to quasi static $\upmu$ is associated with net FM moment arising due to spontaneous spin canting in otherwise AFM. 

 \subsection{Quasi Static Remanence and Piezomagnetism}

 A general consensus in the literature is PzM is connected with the transition from pure AFM to WFM state in an otherwise AFM and one of the mechanism that leads to the WFM state is associated with DMI \cite{Romanov4}. As mentioned before, the $stress$ induced moments have already been experimentally measured in such WFM systems \cite{Romanov1,Romanov2,Romanov3,Romanov4}.  More importantly, the direction of net FM moment in the WFM phase is seen to coincide with the direction of PzM \cite{Romanov1}. It is also to be recalled that waiting-time  effects and ambiguity with sign ( similar to what is observe in remanence data for $\alpha$-Fe$_2$O$_3$ w.r.t the sign of the applied $H$) have also been observed in the sign of stress induced moments in  WFM/PzM on repeated cooling \cite{Romanov1, Sendonis}. 

	The data presented in Fig.\ref{Figure7} explains the ambiguity with the sign and the robustness of the pinned moments in WFM region.  The presence of quasi static remanence  also shows that once the WFM domains have been formed, guided by the magnetic field from above the magnetic transition temperature, removal  of $H$ (or reversing  its sign) does not make any difference.  The net FM moment arise due to DMI driven canting, their direction can be manipulated only when the $H$ is applied from above T$_N$.  It is also well known that magnetization reversal in piezomoments would require the reversal of WFM sublattice  which is energetically unfavorable \cite{Romanov3}.  In remanence measurements, this phenomenon is manifested in the form of  ultra slow magnetization relaxation (and consequently the quasi static remanence)  as observed here.  These data presented in Fig.\ref{Figure2} to Fig.\ref{Figure7}  connect WFM and  quasi static remanence. These data also further confirm that WFM phase is intimately related with the onset of transverse PzM in rhombohedral AFM. 
	
	We emphasize  that the remanence data shown here not only bears a striking  similarity with experimentally measured stress induced moments  but also reveals features  which are not obvious in  routine in-field magnetization data.  Thus it appears that the remanence measurements  capture the essential physics of  DMI driven WFM  better than routine M vs T or M vs $H$ and the onset of quasi static remanence can be taken as footprints of  WFM and PzM.

From present data it also appears that ultra-slow magnetization dynamics and its unusual magnetic field dependence  arises from the WFM and such systems are potential PzM.  The magnitude of the WFM/PzM is further related to lattice parameters, especially $c/a$ ratio in all these rhombohedral systems.   A systematic study of such canonical WFM/PzM such as presented here, points towards the footprints of this phenomenon by simple magnetization measurements. It is to be emphasized that the system considered here are AFM with WFM trait. These are not frustrated AFM or a disordered glassy system / spin glass in conventional sense, which can exhibit slow relaxation for various other reasons. Therefore it is very interesting to observe ultra-slow relaxation in a completely ordered system in which these features are correlated with DMI/SOC. 

From our data, it can be concluded that for micro-cubes of MnCO$_3$ and nano-cubes and single crystal  of $\alpha$-Fe$_2$O$_3$,  the presence of ultra-slow magnetization dynamics is associated with intrinsic WFM. The temperature variation of remanence data on nano-cubes (Fig.\ref{Figure4}a)  and single crystal of $\alpha$-Fe$_2$O$_3$,(Fig.\ref{Figure5}c) especially bring out that the magnitude of quasi static remanence can be significantly tunes by nanoscaling, as also has been observed earlier \cite{Ashna1, Ashna2}. For FeCO$_3$, data is not sufficient to conclude whether effect is intrinsic or it is arising from the size reduction, as the sample comprises of 5-10 nm particles of FeCO$_3$. In such cases, the strain in lattice parameters can also stabilize the WFM phase \cite{Binek,Ashna1,Ashna2}, however microscopic measurements are needed to confirm the presence of DMI driven canting.  It is to be noted that is relatively hard to stabilize FeCO$_3$ in the form of macroscopic crystallites for ruling out size effects. However, we are in the process of exploring systematic size effects in FeCO$_3$. We also assert that for systems which are isostructural AFM with $\alpha$-Fe$_2$O$_3$, such as Cr$_2$O$_3$(which is definitely not a symmetry allowed PzM) and FeCO$_3$ (for which there are conflicting reports in the literature) the strain in the lattice parameter arising from size effects is likely to stabilize the WFM/PzM phase \cite{Ashna1,Ashna2}.

 \section{Conclusion}
  In conclusion, we explore two rhombohedral antiferromagnets that are weak ferromagnets and observe an ultra-slow magnetization dynamics and associated with this, a very robust magnetization pinning with unusual magnetic field dependence.  These features are intimately related to the weak ferromagnetism arising from spin canting.  This spin canting is associated with DMI for the rhombohedral antiferromagnets discussed here. Whether qualitatively similar feature can be observed in other WFM , in which spins are canted but the origin in not DMI driven, is yet to be explored.  From present set of data, it is confirmed that the quasi static remanence and its unique  magnetic  field dependence can be taken as foot prints of WFM/PZM sysyems.This feature is intrinsic in nature and the slow relaxation observed here does not relate with magnetization pinning arising from the glassy phase, magnetocrystalline anisotropy or routine exchange bias. The DMI in WFM phase is clearly connected with the possibility of stress induced moments or piezomagnetism. Finally, piezomagnetism, though not as widely explored or utilized, say as piezoelectricity, can have a variety of applications including those related to FM/AFM interfaces, in which the FM moment can be pinned by a PzM, and the effect should be tunable by stress alone.  
  
\section{Acknowledgments} 
Authors thank Sunil Nair (IISER Pune) for SQUID magnetization measurements. AB acknowledges Department of Science and Technology (DST), India for funding support through a Ramanujan Grant and the DST Nanomission Thematic Unit Program. SWC is funded by the Gordon and Betty Moore Foundation’s EPiQS Initiative through Grant GBMF4413 to the Rutgers Center for Emergent Materials. Authors thank the DST and Saha Institute of Nuclear Physics, India for facilitating the experiments at the Indian Beamline, Photon Factory, KEK, Japan.  

\bibliography{Bibliography}
\end{document}